\documentclass[showpacs,prb,twocolumn,floatfix]{revtex4}

\usepackage{graphicx,float,amsmath}

\begin{document}

\title{Imaging ambipolar diffusion of photocarriers in GaAs thin films}
\author{D. Paget, F. Cadiz, A.C.H. Rowe}
\affiliation{Physique de la mati\`ere condens\'ee, Ecole Polytechnique, CNRS, 91128
Palaiseau, France}
\author{F. Moreau}
\affiliation{Physique des interfaces et des couches minces, Ecole Polytechnique, CNRS, 91128
Palaiseau, France}
\author{S. Arscott, E. Peytavit}
\affiliation{Institut d'Electronique, de Micro\'electronique et de Nanotechnologie
(IEMN), CNRS UMR8520, Avenue Poincar\'e, Cit\'e Scientifique, 59652
Villeneuve d'Ascq, France}

\begin{abstract}
Images of the steady-state luminescence of passivated GaAs self-standing films under excitation by a tightly-focussed laser are analyzed as a function of light excitation power. While unipolar diffusion of photoelectrons is dominant at very low light excitation power, an increased power results in a decrease of the diffusion constant near the center of the image due to the onset of ambipolar diffusion. The results are in agreement with a numerical solution of the diffusion equations and with a physical analysis of the luminescence intensity at the centre of the image, which permits the determination of the ambipolar diffusion constant as a function of electron concentration.
\end{abstract}

\maketitle

\section{Introduction}

Ambipolar diffusion is the term used to describe the diffusion of electrons and holes in semiconductors when their respective concentrations are such that the electrostatic coupling between the two populations can no longer be neglected. From a practical viewpoint, this phenomenon must be accounted for when designing any bipolar device. After the initial work on electrostatic coupling between electrons and holes \cite{smith78}, significant theoretical and experimental work has been published on ambipolar diffusion in bulk materials \cite{meyer80,young82} as well as in heterostructures \cite{zarem89, gulden91}. The majority of recent studies consider undoped material so that the ambipolar diffusion constant is only related to hole diffusion \cite{ruzicka10,zhao09,zhao08} or to excitonic transport \cite{chao99}. The dependence of the ambipolar diffusion constant, $D_a = (D_n\sigma_p+D_p\sigma_n)/(\sigma_p+\sigma_n)$, on the unipolar diffusion constant $D_n$($D_p$) of electrons (holes) and of their partial conductivities $\sigma_n$($\sigma_p$) has never been detailed experimentally. Furthermore, the effect of the electric field induced by spatial separation of electrons and holes has never been evaluated precisely. 

\begin{figure}[t]
\includegraphics[clip,width=8 cm] {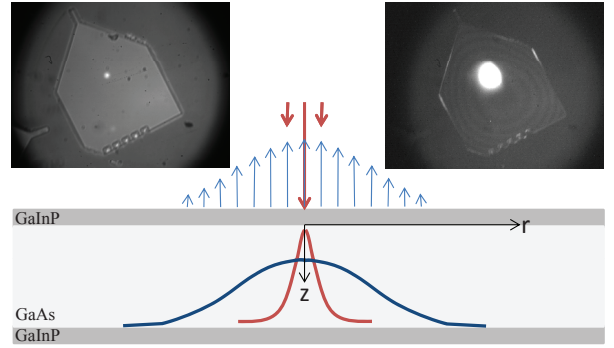}
\caption{Principle of the experiment: A thin, self-supported (3$\mu$m) GaAs sample is excited by tightly-focused above bandgap light (red arrows and top, left inset). An image of the bandgap emission is monitored (blue arrows and top, right inset). Since the surface recombination is quenched by thin GaInP films above and below the GaAs, this image reveals the diffusion of carriers within the GaAs.}
\label{fig1}
\end{figure}

Here we present an optical investigation of ambipolar diffusion of photoexcited carriers in a thin slab of \textit{p}$^+$ GaAs (3 $\mu$m thickness) passivated on both sides by 50 nm thick GaInP layers (see Fig. \ref{fig1}). The sample is excited at its center by a tightly-focused laser along the $z$ direction such that steady-state imaging of the luminescence intensity enables us to monitor the diffusion profile of minority carriers \cite{favorskiy10}.  The resulting profiles are interpreted using two distinct and complementary approaches: i) a numerical resolution of the coupled diffusion equations for electrons and holes, and ii) a simple qualitative estimate of the electron concentration at the center which yields the power dependence of the luminescence thereby permitting $D_a$ to be evaluated as a function of photoelectron concentration. 

\section{Ambipolar diffusion of carriers in a thin semiconducting slab}

\subsection{Coupled diffusion equations}

In photo-excited \textit{p}$^+$ GaAs the drift-diffusion equations for electrons and holes are \begin{equation} \label{electrons} \frac{\partial n}{\partial t} = g - K(N_A+\delta p)n+\vec{\nabla}\cdot\left[\mu_n n \vec{E}+D_n\vec{\nabla}n\right] \end{equation} and \begin{equation} \label{holes} \frac{\partial \delta p}{\partial t} = g - K(N_A+\delta p)n+\vec{\nabla}\cdot\left[-\mu_p(N_A+\delta p) \vec{E}+D_p\vec{\nabla}\delta p\right] \end{equation}
where $\delta p$ is the concentration of photogenerated holes and $N_A$ is the concentration of acceptors which (in the following discussion) will be assumed to be fully ionized. $K$ is the bimolecular electron-hole recombination coefficient and $\mu_n$ and $\mu_p$ are the electron and hole mobilities respectively. Non-radiative bulk recombination is neglected for the purposes of this discussion. The terms involving the electric field ($\vec{E}$) in Eqs. \ref{electrons} and \ref{holes} are responsible for the electrostatic coupling between electrons and holes. In this case $\vec{E}$ is the internal electric field resulting from the spatial distribution of electrons and holes. It is given by the Poisson equation \begin{equation} \label{poisson} \vec{\nabla}\cdot\vec{E} = \frac{q}{\epsilon}(\delta p - n) \end{equation} where $\epsilon$ is the permittivity and $q$ the absolute electronic charge. By equating Eqs. \ref{electrons} and \ref{holes} in steady-state, an independent expression for the electric field in terms of the diffusion constants and concentration gradients can be obtained. Using this one may re-write the drift-diffusion equation for electrons in the form: \begin{equation} \label{elecs} 0 = g - K(N_A+\delta p)n+\vec{\nabla}\cdot\left[D_a \vec{\nabla}n -D_a^\prime\vec{\nabla}(n-\delta p)\right] \end{equation} where \begin{equation} \label{Da} D_a = \frac{D_n \mu_p (N_A+\delta p)+D_p \mu_n n}{\mu_n n + \mu_p (N_A+\delta p)} \end{equation} is the usual value of the ambipolar diffusion constant and $D_a^\prime = D_p \mu_n n/(\mu_n n + \mu_p (N_A + \delta p))$ gives the magnitude of the correction due to the local departure $(\delta p - n)$ from charge neutrality. The spatial distributions of electron and hole concentrations are finally calculated using Eq. \ref{holes}, Eq. \ref{elecs} and Eq. \ref{poisson}.

Since ambipolar diffusion will be evaluated by varying the incident light power and hence the photoelectron concentration, the effect of Fermi blockade on the diffusion constants should also be accounted for. In this case the diffusion constant depends on the electron concentration via the position of the quasi-Fermi level, $E_{Fe}$, when the photo-electron concentration becomes comparable with the effective density of states of the conduction band (i.e. when the electron gas becomes weakly degenerate). The electron diffusion constant is then written \begin{equation} \label{blockade} D_n = 2D_n^0 \frac{F_{1/2}(E_{Fe}/k_BT)}{F_{-1/2}(E_{Fe}/k_BT)} \end{equation}
where $F_n(\phi) = \int_0^\infty x^n (\exp(x-\phi)+1)^{-1} \mathrm{d}x$ and $D_n^0 = \mu_n k_BT/q$ is the low concentration (non-degenerate) value of the electron diffusion constant. Here  $k_B$ is the Boltzmann constant and $T$ is the temperature. The Fermi energy is related to the electron concentration in the conduction band by $n = \int_0^\infty x^n \rho(\phi)(\exp(\phi-E_{Fe})+1)^{-1} \mathrm{d}\phi$ where $\rho(\phi)$ is the density of states in the conduction band at energy $\phi$. 

\begin{figure}[t]
\includegraphics[clip,width=8 cm] {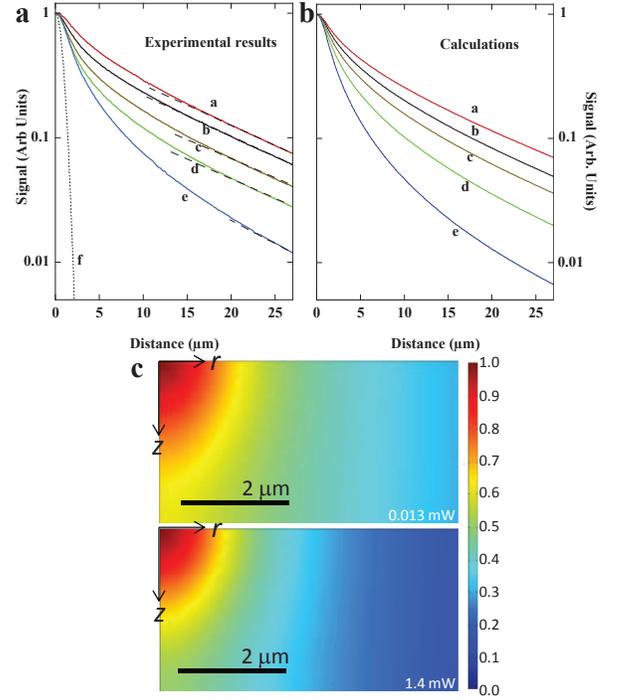}
\caption{(a) The normalized luminescence cross section for a light excitation power of 0.013 mW, 0.096 mW, 0.24 mW, 0.49 mW, and 1.4 mW (curves a to e respectively). Curve f is the laser profile. (b) Self consistent calculations under the same conditions. In both cases the shape of the profile at low power and large $r$ yields a diffusion length of 21.5 $\mu$m. The effect of an increase of light intensity is to decrease the diffusion length near $r = 0$ due to ambipolar diffusion, while the slope of the logarithmic plot remains practically unchanged at large $r$ where $u = n/N_A$ is small. (c) Calculated spatial dependence of the normalized electronic concentration. The top frame shows the low power case (curve a of the right panel), while the bottom frame shows the high power case (curve e of the right panel).}
\label{fig2}
\end{figure}

\subsection{Power dependence of the luminescence at the center.} 

It is shown here that simple estimates of the electronic concentration at the center of the image ($r$ = 0) can be used to qualitatively investigate the unipolar and ambipolar diffusion regimes. For relatively low powers it is assumed that the effect of degeneracy on the Einstein relation is weak, so that the low concentration value $D_n^0$ of the diffusion constant can be used. As will be verified \textit{a posteriori}, at the center ($r = 0$) it is reasonable to assume charge neutrality ($n=\delta p$) so that Eq. \ref{elecs} only contains the generation, recombination and ambipolar diffusion terms. Secondly, since the diffusion length is larger than the Gaussian width $\sigma$ of the laser spot, diffusion dominates bulk and surface recombination and is thus the determining factor for the steady-state photoelectron concentration at the center. Assuming that diffusion parallel to the surface can be characterized by a rate $\tau_d$, one has \begin{equation} \label{sigma} \sigma^2 = D_a \tau_d \xi^{-1} = D_a \tau_d^* \end{equation} where $\xi$  is a numerical factor close to unity. At low power, using the value of the unipolar diffusion constant, one finds $\tau_d^* \approx 6\times10^{-11}$ s, i.e. about three orders of magnitude smaller than the typical photoelectron lifetime $\tau$ of \textit{p}$^+$ GaAs \cite{lowney91}.

\begin{figure}[t]
\includegraphics[clip,width=8 cm] {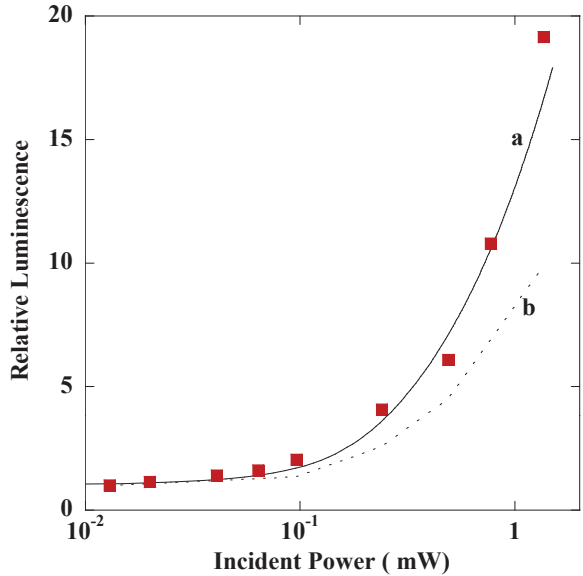}
\caption{Luminescence at $r = 0$, normalized with respect to the light excitation power as a function of incident power (red squares). Curve a is the adjustment calculated using Eq. \ref{solution} and curve b is the result of the ab-initio calculation.}
\label{fig3}
\end{figure}

After generation over a characteristic depth $1/\alpha$, where $\alpha$ is the absorption coefficient, the photoelectrons and holes undergo lateral diffusion as well as diffusion along $z$. Since the latter does not change the luminescence intensity, the one dimensional diffusion along $z$ can be treated independently of the lateral, two dimensional diffusion. This permits the computation of quantities averaged along $z$ over the thickness $d$ of the sample. The average rate of creation of the photoelectron concentration is given by \begin{equation} \label{g} g = P(1-R)\zeta/(h\nu\pi\sigma^2 d) = \zeta g^* \end{equation} where $R$ is the reflectivity of the sample surface and $\zeta$ is a numerical factor close to unity. Considering $n$ to be homogeneous as a function $z$, its value is given by $n = \eta g^* \tau_d^*$ where $\eta = \zeta \xi$ summarizes the above approximations. Using Eqs. \ref{Da}, \ref{sigma} and \ref{g}, one finds that $n$ does not depend on the size of the laser spot $\sigma$. It is the solution of the second degree equation \begin{equation} \label{u} 2u^2+[1-p(1+\beta)]u - p = 0 \end{equation} where $\beta=\mu_n/\mu_p$ and the reduced values of concentration and power are $u = n/N_A$ and \begin{equation} \label{power} p = \frac{\eta P (1-R)}{\pi h \nu d D_n^0 N_A}=\frac{P}{P^*}. \end{equation}
Here $P^* = \pi h \nu d D_n^0 N_A/[\eta(1-R)]$ is a power. The solution of this equation is \begin{equation} \label{solution} 4u = p(1+\beta)-1+\sqrt{8p+[1-p(1+\beta)]^2} \end{equation}
As seen from the above approximations, and using Eq. \ref{electrons}, the luminescence at the center is proportional to $u(1+u)$. 

\section{Experimental}

The samples are \textit{p}$^+$ beryllium-doped GaAs thin films ($N_A \approx 10^{17}$ cm$^{-3}$) of thickness 3 $\mu$m assembled onto SiC substrates \cite{arscott10}. Reduction of the surface recombination velocity is provided by 50 nm thick layers of Ga$_{0.51}$In$_{0.49}$P deposited on each face of the film. The samples are excited by a tightly-focussed laser beam of energy 1.59 eV in a modified Nikon Optiphot 70 microscope. The laser profile, shown in curve f of Fig. \ref{fig1}, is close to a Gaussian profile $\exp[-r^2/\sigma^2]$ with $\sigma \approx 0.93 \mu$m. The luminescence cross sections were recorded using an appropriate filter in order to filter out the excitation wavelength. Curve a was taken using a very low excitation power (13 $\mu$W). The spatial extent of these profiles is much larger than that of the laser, thus revealing electron diffusion in the film. At this very low power, diffusion is assumed to be unipolar, and the whole cross section is interpreted using a diffusion length $L = 21.3 \mu$m \cite{nelson78}, and is nearly exponential for $r > 12 \mu$m. Shown in curves b, c, d and e of Fig. \ref{fig2} are the spatial dependences of the normalized cross sections for increasingly high powers up to 1.4 mW, above which the luminescence spectrum reveals a heating of both the electron gas and the lattice. Curves b, c, d, and to some extent e show little change of the profile slope far from the center where the photoelectron concentration is small, thus revealing that the electron diffusion constant at large radii is close to its unipolar value. On the other hand at small radii, the slope strongly increases indicating a reduction of diffusion constant to its ambipolar value given by Eq. \ref{Da} for large photoelectron concentrations. 

Shown in Fig. \ref{fig3} is the luminescence magnitude at the center, $I_{PL}(0)$ normalized to the incident power and to a value of 1 at the lowest power value. This signal is close to unity up to about 0.1 mW and reaches values larger than 20 for the maximum excitation power. The relative excess of carriers at the center is consistent with the decrease of diffusion constant due to the progressive onset of ambipolar diffusion. 

\section{Interpretation}
\label{interp}

In order to interpret the above results, the photoelectron lifetime $\tau$ was first measured using time-resolved microwave conductivity \cite{brenot01} and found equal to 30.7 ns. The good correspondence with the radiative recombination time for the nominal doping level \cite{nelson78} is further proof that nonradiative surface and bulk recombination processes are negligible. Since $L$ is known at very low excitation power (curve a of Fig. 2), the value of the unipolar diffusion constant can be estimated to be  $D_n^0 \approx 150$ cm$^2/$s. Finally, as found from the literature \cite{lowney91} one has $\beta \approx 10$ so that the relevant quantities describing charge diffusion are known. 
 
\subsection{Electronic concentration and luminescence intensity at $r = 0$.}
\label{cent}

The normalized luminescence intensity shown in Fig. \ref{fig3} is given by \begin{equation} \label{Ipl} I_{PL}(0) = \frac{P_0(1+u)u}{P(1+u_0)u_0} \end{equation} where u is the normalized electronic concentration defined in Sec. II, and $P_0$ is the smallest experimental power value, corresponding to $u = u_0$. Shown in Fig. \ref{fig3} is the calculated power dependence of $I_{PL}(0)$, using $P^* = $2 mW. Very good agreement is then obtained using Eq. \ref{solution}, $N_A = 1\times10^{17}$ cm$^{-3}$ and a power-independent value, close to unity, of $\eta \approx 1.25$. This justifies the main physical, but not completely trivial, approximations made for obtaining the expression for $n$. 

The calculated power dependences of the reduced values of the ambipolar diffusion constant $D_a/D_n^0$ and of the luminescence intensity $u(1+u)$ are shown versus $u$ in curves a, c, and d of Fig. \ref{fig4}. Switching from the unipolar to the ambipolar regime is revealed by the decrease in the diffusion constant. For the maximum value of $u$ one finds $D_a/D_n^0 \approx 0.2$. This result is in agreement with Eq. \ref{Da}, which gives $D_a \approx 2D_p \approx 2\beta^{-1}D_n^0$ in the limit where $n \gg N_A$. It is also seen that $u$ increases faster than the light power and that its value at maximum power is of the order of $3N_A$. The power dependence of the luminescence intensity starts to differ from that of the electron concentration for $P \approx 0.1$ mW.

Since the electron concentration at high power is comparable with the intrinsic density of states in the conduction band, the effect of the concentration dependence of the electron diffusion constant, described by Eq. \ref{blockade}, needs to be evaluated. To first order, taking $D_n$ of the form $D_n = D_n^0(1+n/n_0)$, one finds $n_0 \approx 1.2\times10^{18}$ cm$^{-3}$. Eq. \ref{u} becomes a third degree equation including the parameter $N_A/n_0$. Shown in curve b of Fig. \ref{fig4} is the resulting power dependence of the ambipolar diffusion constant. $D_n$ only differs from $D_n^0$ (curve a) for powers larger than about 0.3 mW. For the maximum power ($P =$ 1.4 mW) the increase in $D_n$ gives a value of $u$ slightly smaller than that shown in curve c and corresponds to $(D_n - D_n^0)/D_n^0 \approx 12$ \%. Given that this marginal increase is not unambiguously evident from the data, it is reasonable to take a concentration-independent electronic diffusion constant. 

\subsection{Luminescence profiles}

In a separate, complementary approach, the coupled equations, Eq. \ref{holes}, Eq. \ref{poisson} and Eq. \ref{elecs} were solved self-consistently using a commercial finite element package. This yields the electronic concentration and the photoluminescence intensity at all positions within the sample. The bottom panel of Fig. \ref{fig2} shows the normalized maps of electronic concentrations near the center for the smallest and for the largest power. It is first verified that at $r = 0$ the relative variation of $n$ as a function $z$ is of the order of 40\% at small power and of 50\% at large power. This \textit{a posteriori} justifies the assumption of homogeneous concentration as a function of depth taken Sec. IIB.  Furthermore since $L \gg d$, the variation of $n(r,z)$ as a function of $z$ is quite weak in both cases as soon as $r$ is comparable with the thickness (note that the horizontal scale in Fig. \ref{fig2}c is much smaller than in Fig. \ref{fig2}a). For numerical calculation of the photoluminescence profiles as a function of $r$ it is therefore not a bad approximation to take $z = d/2$. 

The calculated luminescence profiles are shown in the right panel of Fig. \ref{fig2} after normalization at $r = 0$ for $N_A = 2.5\times10^{17}$ cm$^{-3}$. The overall behavior of the experimental profiles is correctly interpreted by the model described above, although slight differences between the ab-initio calculations and the experimental results are apparent. This is most evident at high power where the shape of the profile depends very sensitively on the reduced concentration, $u$ (i.e. on the exact doping density and on the incident power). Any small variation in $N_A$ (whose exact value is not known) or in the incident power results in a large relative variation of the luminescence intensity for large $r$. For example, the use of $N_A = 1\times10^{17}$ yields an $r$ dependence of the normalized luminescence profile that is far too strong. Curve b of Fig. \ref{fig3} shows $I_{PL}(0)$. In the case of the ab-initio calculation, the ratio is calculated after integration over the whole thickness of the sample and over a lateral radius of the order of that of the excitation spot. Once again, although the qualitative shapes of the calculated and experimental curves are in reasonable agreement, there are quantitative differences between the curves. As above, this is particularly so at high power where the luminescence intensity depends sensitively on $u$. Undoubtedly better agreement could be obtained by varying several parameter values ($N_A$, $\beta$, etc\dots) but doing so is tedious and not particularly revealing from a physical point of view. It is also possible that the slight difference is due to photon recycling which could yield a luminescence profile somewhat larger than that due to carrier diffusion alone \cite{asbeck77,kuriyama77}.

\begin{figure}[t]
\includegraphics[clip,width=8 cm] {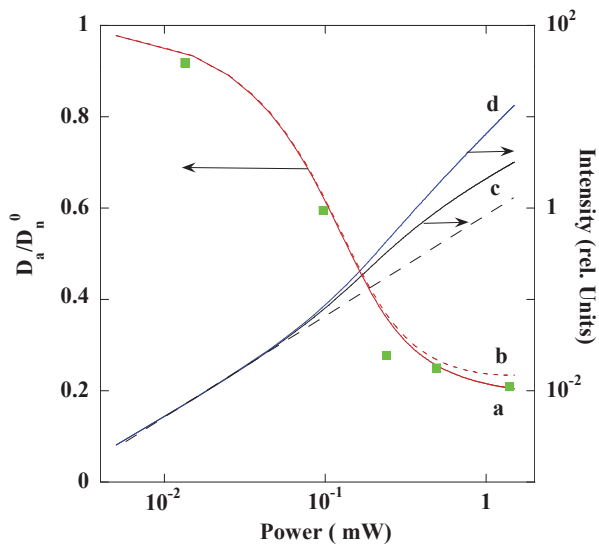}
\caption{The calculated dependence of the diffusion coefficient normalized to its unipolar value as a function of light excitation power. Curve a(b) corresponds to the result calculated without(with) Fermi blockade taken into account. Also shown are to dependences of the electronic concentration $u$ normalized to the acceptor concentration (curve c) and of $u(1+u)$ (curve d) which is proportional to the luminescence intensity. The results of the full numerical calculation (green squares) are in excellent agreement with the simplified analytical approach. Shown for comparison is a straight, dotted line of slope unity.}
\label{fig4}
\end{figure}

\begin{figure}[t]
\includegraphics[clip,width=8 cm] {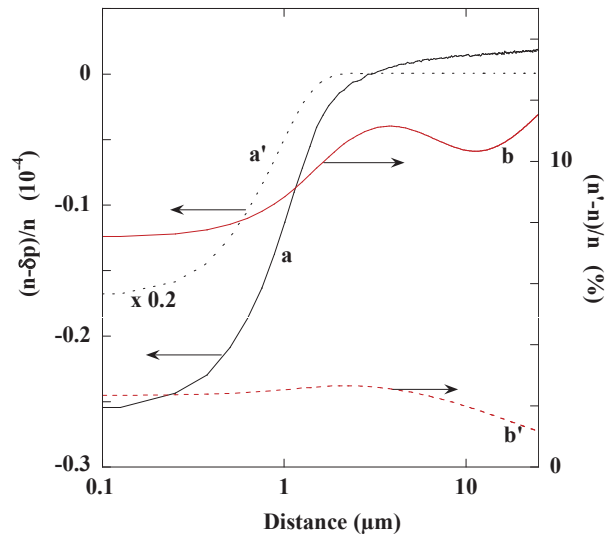}
\caption{Estimation of the internal electrical field and the validity of the local neutrality approximation for an incident powers of 0.013 mW (curves a' and b') and 1.4 mW (curves a and b). Curves a and a' show the spatial distribution of the relative difference between electron and hole concentration. Curves b and b' show the relative difference  between the spatial distribution of electrons obtained when $n = n^\prime = \delta p$ in Eq. \ref{elecs} and when the electric field is accounted for.}
\label{fig5}
\end{figure}

One advantage of the \textit{ab-initio} calculation is that it can be used to evaluate the assumption of local charge neutrality (i.e. $n = \delta p$) that is made in all discussions of ambipolar transport \cite{smith78}. Fig. \ref{fig5} shows the spatial distribution of the relative difference $(n-\delta p)/n$ at $z = d/2$ for the lowest (curve a') and highest (curve a) incident powers. In both cases there is an excess of holes near $r=0$ and a compensating excess of electrons at a distance larger than 3-4 $\mu$m. As expected, the relative excess of holes at the center is larger at low powers where ambipolar diffusion is absent. In the presence of ambipolar diffusion, electrons and holes have a tendency to diffuse together and the relative difference drops by a factor of 10. Since the permittivity $\epsilon$ in Eq. \ref{poisson} is small, these observations do not necessarily imply that the term proportional to $\vec{E}$ is negligible. In order to validate the assumption of local charge neutrality, the electronic concentration $n^\prime$ obtained when neglecting the last term of Eq. \ref{elecs} is calculated. Shown in curves b and b' of Fig. \ref{fig5} is the relative value $(n^\prime-n)/n$ for the highest and lowest powers respectively. Unsurprisingly, the term proportional to $\vec{E}$ is more important at higher power , although at worst, assuming $n = \delta p$ introduces an error of the order of 10\% into the resulting concentration profiles. More importantly, the error is smallest at $r = 0$, indicating that the simplifying assumptions used above to analyze the luminescence intensity at the center are reasonable. This is confirmed by the excellent agreement obtained between the exact numerical and approximate analytic calculations of $D_a/D_n^0$ at $r=0$ shown in Fig. \ref{fig4}.

\section{Conclusion}

Imaging of the luminescence profile created by a highly focused excitation and emitted by a 3$\mu$m thick \textit{p}$^+$ GaAs clearly reveals ambipolar diffusion as the excitation power is increased. The switching from unipolar to ambipolar diffusion of photocarriers is investigated as a function of electron concentration and the results are analyzed using a numerical resolution of the coupled electron and hole diffusion equations, as well as the Poisson equation. It is found that the effect of the electric field induced by ambipolar diffusion can be significant away from the center. In contrast, this effect is reduced near the center so that a simple calculation of the power dependence of the luminescence intensity can be performed. The results are interpreted using a single parameter, defined in Eq. \ref{power} as a power $P^*$, which depends on acceptor concentration, slab thickness and unipolar electron diffusion constant. The experimental results at the center are in very good agreement with the predictions of this model, using a reasonable value of $P^*$. 

\bibliographystyle{apsrev}
\bibliography{bibroweambi}

\end{document}